\documentclass[%
 reprint,
 superscriptaddress,
 amsmath,amssymb,
 aps,
]{revtex4-2}

\usepackage{graphicx}
\usepackage{dcolumn}
\usepackage{bm}
\usepackage{hyperref}
\usepackage{physics}
\usepackage{algorithm}
\usepackage{algorithmicx}
\usepackage{algcompatible}
\usepackage[end]{algpseudocode}
\hypersetup{colorlinks=true, linkcolor=blue, citecolor=blue, urlcolor=blue, unicode=true}
\usepackage{xcolor}
\usepackage{float}
\usepackage{ulem}

\begin{document}

\preprint{APS/123-QED}

\title{Data-enhanced self-learning projective quantum Monte Carlo \\ with Rydberg-atom
measurements}

\author{V. Imchitr}
 \email{vimchitr@perimeterinstitute.ca}
  \affiliation{%
  Department of Physics and Astronomy, University of Waterloo, 200 University Ave W, Waterloo, ON N2L 3G1, Canada
}%
\affiliation{%
 Perimeter Institute for Theoretical Physics, Waterloo, Ontario N2L 2Y5, Canada
}%

\author{L. Brodoloni}%

\author{S. Pilati}
\affiliation{
{CQM group, School of Science and Technology, Physics Division, Università di Camerino, 62032 Camerino, Italy}
}%
\affiliation{{INFN, Sezione di Perugia, I-06123 Perugia, Italy}}

\author{E. M. Inack}
\affiliation{%
 Perimeter Institute for Theoretical Physics, Waterloo, Ontario N2L 2Y5, Canada
}%
  \affiliation{%
 Department of Physics and Astronomy, University of Waterloo, 200 University Ave W, Waterloo, ON N2L 3G1, Canada
}%
\affiliation{yiyaniQ, Toronto, ON, Canada}

\date{\today}

\begin{abstract}

We introduce a data-enhanced self-learning projective quantum Monte Carlo
(PQMC) method that incorporates projective measurements from programmable
quantum simulators into the guiding wavefunction of a ground-state
projection. We demonstrate the approach for a two-dimensional Rydberg-atom
array using restricted Boltzmann machine (RBM) and recurrent neural network
(RNN) guiding wavefunctions. Experimental pretraining reduces
finite-population effects and accelerates self-learning convergence, allowing unbiased estimation of
ground-state properties to be obtained with fewer walkers and fewer self-learning iterations. The improvement
is particularly pronounced for the RBM, while the more accurate RNN guiding wavefunction exhibits a smaller, though still observable, benefit from pretraining. Simulations across different system sizes suggest that the relative advantage of experimental pretraining may increase with system size. Improvements are also observed for the checkerboard order parameter, despite deviations of the experimental measurements from the equilibrium numerical results. Finally, finite-size
scaling with self-learning PQMC without experimental pretraining yields the
critical point and critical exponents of the
disordered-to-checkerboard transition in agreement with previous numerical
studies. Our results show that limited and imperfect experimental measurements can provide an effective initialization for self-learning PQMC, while imaginary-time
projection combined with subsequent unsupervised learning progressively refines the guiding wavefunction toward the ground state.

\end{abstract}

\maketitle
\section{\label{sec:intro}Introduction}
An accurate theoretical description of generic strongly-correlated quantum many-body systems is fundamentally hindered by the exponential growth of the Hilbert space with increasing system size, posing severe scalability challenges for many classical numerical methods. Stochastic approaches partly circumvent this bottleneck by sampling relevant many-body configurations without explicitly exploring the entire Hilbert space. Among these, projective quantum Monte Carlo (PQMC) provides a rigorous framework for accessing ground-state properties by stochastically sampling the imaginary-time evolution of a trial wavefunction using a population of random walkers~\cite{Becca17, foulkes01, teivedi90, buonaura98}. For stoquastic Hamiltonians, PQMC is free of the negative-sign problem and converges to the exact ground state in the long imaginary time and large walker population limits. In practice, the walker population required to reach convergence can be greatly reduced by introducing importance sampling, a procedure in which an ansatz referred to as a \textit{guiding wavefunction} is used to bias the dynamics towards the most relevant regions of the Hilbert space. The efficiency of PQMC therefore heavily relies on the quality of the guiding wavefunction~\cite{nemec10, boninsegni12, inack18-1, perez2026unbiaseddiffusionmontecarlo} and, in particular, on how accurately it approximates ground-state wavefunctions.

Recent developments in neural quantum states (NQS) offer a natural route to constructing such accurate guiding wavefunctions, owing to their ability to efficiently represent nontrivial correlations in quantum matter~\cite{androsiuk93, carrasquilla17, chen24, lage2024, rende25}.  Generative models such as restricted Boltzmann machines (RBMs) \cite{carleo17, choo17, nomura17, melko19} and recurrent neural networks (RNNs)~\cite{hibat-allah20, hibatallah22, mcnaughtonarxiv25} have been successfully employed as variational ansätze in variational Monte Carlo and for quantum-state reconstruction from measurement data. Their integration into PQMC as guiding wavefunctions has been shown to substantially reduce the walker population required for convergence and improve the accuracy of ground-state estimates~\cite{inack18, ren23, wilsonarxiv}. Furthermore, the self-learning protocol of PQMC~\cite{pilati19, wah2026autoregressiveprojectivequantummonte} can exploit the walker population itself to iteratively train the NQS from scratch via unsupervised learning, thereby progressively improving the guiding wavefunction during the simulation. This approach retains the asymptotic exactness of PQMC while simultaneously yielding an accurate NQS representation of the ground state as a byproduct~\cite{pilati19, pilati20, brodoloni24, wah2026autoregressiveprojectivequantummonte}.

In parallel with these classical algorithmic developments, programmable quantum simulators based on neutral-atom arrays have emerged as a powerful platform for studying strongly correlated quantum matter~\cite{adams20, saffman16, morgado21, browaeys20, henriet20}. In these platforms, atoms trapped in reconfigurable optical-tweezer arrays encode qubits in superpositions of their ground and highly excited Rydberg states, with entanglement generated through long-range Rydberg interactions~\cite{lukin01, urban09, jaksch00}. This combination enables the simulation of rich many-body phases of matter~\cite{kalinovski22, ebadi21, samajdar21} and has also found applications in combinatorial optimization~\cite{scholl21, schauss17, chen23, ebadi22}. Despite limitations from finite coherence times, data-acquisition rates, and state-preparation-and-measurement errors~\cite{bluvstein22}, projective measurements from these devices can provide valuable information to classical algorithms. Combined with machine learning, such data have been used for quantum state reconstruction and data-enhanced variational Monte Carlo~\cite{torlai19, moss24}. Furthermore, recent studies have shown that NQS trained on data from quantum devices can accelerate classical quantum Monte Carlo techniques, pointing toward hybrid approaches in which quantum-generated data enhance classical many-body methods via deep learning \cite{jiang26, provazza26}.
Motivated by this hybrid quantum-classical paradigm, we build on the long-standing complementarity between variational and projective approaches. Traditionally, projective Monte Carlo methods have been used to systematically improve approximate variational states, as imaginary-time projection filters out excited-state components remaining in the variational ansatz. This suggests that an NQS pretrained on imperfect quantum-generated data need not itself accurately represent the ground state to provide a useful starting point for PQMC. Self-learning PQMC iteratively refines the initial data-driven NQS toward the projected ground-state distribution using random walkers sampled during projection.

Building on these observations, we integrate experimental data from a programmable quantum simulator into self-learning PQMC simulations of a two-dimensional Rydberg-atom system on a square lattice. The resulting framework, which we dub \textit{data-enhanced self-learning PQMC}, is illustrated in Fig.~\ref{fig:workflow} and Algorithm~\ref{alg:data_slpqmc}. We employ both RBM and RNN guiding wavefunctions to assess how the quality of the neural representation affects the efficiency of the projection. We find that RNNs provide more accurate guiding wavefunctions than RBMs and consequently more efficient PQMC simulations. Pretraining the guiding wavefunctions on projective measurement data further reduces the walker population required to obtain accurate ground-state properties and accelerates convergence relative to conventional self-learning PQMC, with the improvement being particularly pronounced for the RBMs. As an independent benchmark of the method, we use self-learning PQMC to extract the critical properties of the disordered-to-checkerboard transition through finite-size scaling, obtaining results consistent with the literature. Taken together, our results show that imperfect quantum-device data can provide a useful initialization for an accurate classical projection method, establishing a hybrid route in which experimental information improves computational efficiency while self-learning PQMC systematically refines the resulting many-body description.

The remainder of this article is organized as follows. In Sec.~\ref{sec:methods}, we introduce the Rydberg Hamiltonian and describe the data-enhanced self-learning PQMC framework and the NQS guiding wavefunctions. In Sec.~\ref{sec:results}, we assess the improvements in walker-population requirements and convergence time enabled by experimental pretraining, and benchmark self-learning PQMC through the study of the disordered-to-checkerboard phase transition. Finally, in Sec.~\ref{sec:level3}, we summarize our findings and discuss future perspectives.
\section{\label{sec:methods}METHODOLOGY}
\subsection{Rydberg-atom array}
\label{sec:rydberg}

We consider two-dimensional $L\times L$ square arrays of Rydberg atoms held in optical tweezers, with $N=L^2$ sites. Computational basis states are realized by the atomic ground state $\ket{g}$ and the highly excited Rydberg state $\ket{r}$. The many-body dynamics of the system are governed by the combination of coherent driving and long-range van der Waals interactions between the Rydberg states, resulting in the Rydberg Hamiltonian:
\begin{equation}
\label{eq:hamiltonian}
\hat{H}=\Omega\sum_{i<j}\left(\frac{R_b}{R_{ij}}\right)^6\hat{n}_i\hat{n}_j
-\Delta\sum_i\hat{n}_i+\frac{\Omega}{2}\sum_i\hat{\sigma}_i^x,
\end{equation}
where $\Omega$ is the Rabi frequency, $\Delta$ is the laser detuning, $R_{ij}$ is the distance between atoms $i$ and $j$, and $R_b$ is the blockade radius. The number operator $\hat{n}_i=(\hat{\sigma}_i^z+1)/2=\ket{r}_i\bra{r}_i$ measures the Rydberg excitation density, while $\hat{\sigma}_i^x$ generates the coherent transition between $\ket{g}_i$ and $\ket{r}_i$. Although the Rabi term in Eq.~\eqref{eq:hamiltonian} has a positive sign, a global $\pi$ rotation about the $z$ axis on every site maps $\hat{\sigma}_i^x\to-\hat{\sigma}_i^x$ and renders $\hat{H}$ stoquastic in the computational basis, hence amenable to sign-problem-free quantum Monte Carlo. We work in this rotated frame throughout.

We focus on the disordered-to-checkerboard transition of the $16\times16$ array of $^{87}$Rb atoms as described in Ref.~\cite{ebadi21}, which has been studied extensively~\cite{ebadi21,kalinovski22}. Unless otherwise stated, $R_b/a\approx1.15$ and $\Omega=2\pi\times4.3$ MHz, with $a$ being the lattice spacing. More details on the experimental setup and the dataset used to pretrain the guiding wavefunctions are provided in Appendix~\ref{app:experimental}.

\begin{figure}[t]
  \centering
  \includegraphics[width=\columnwidth]{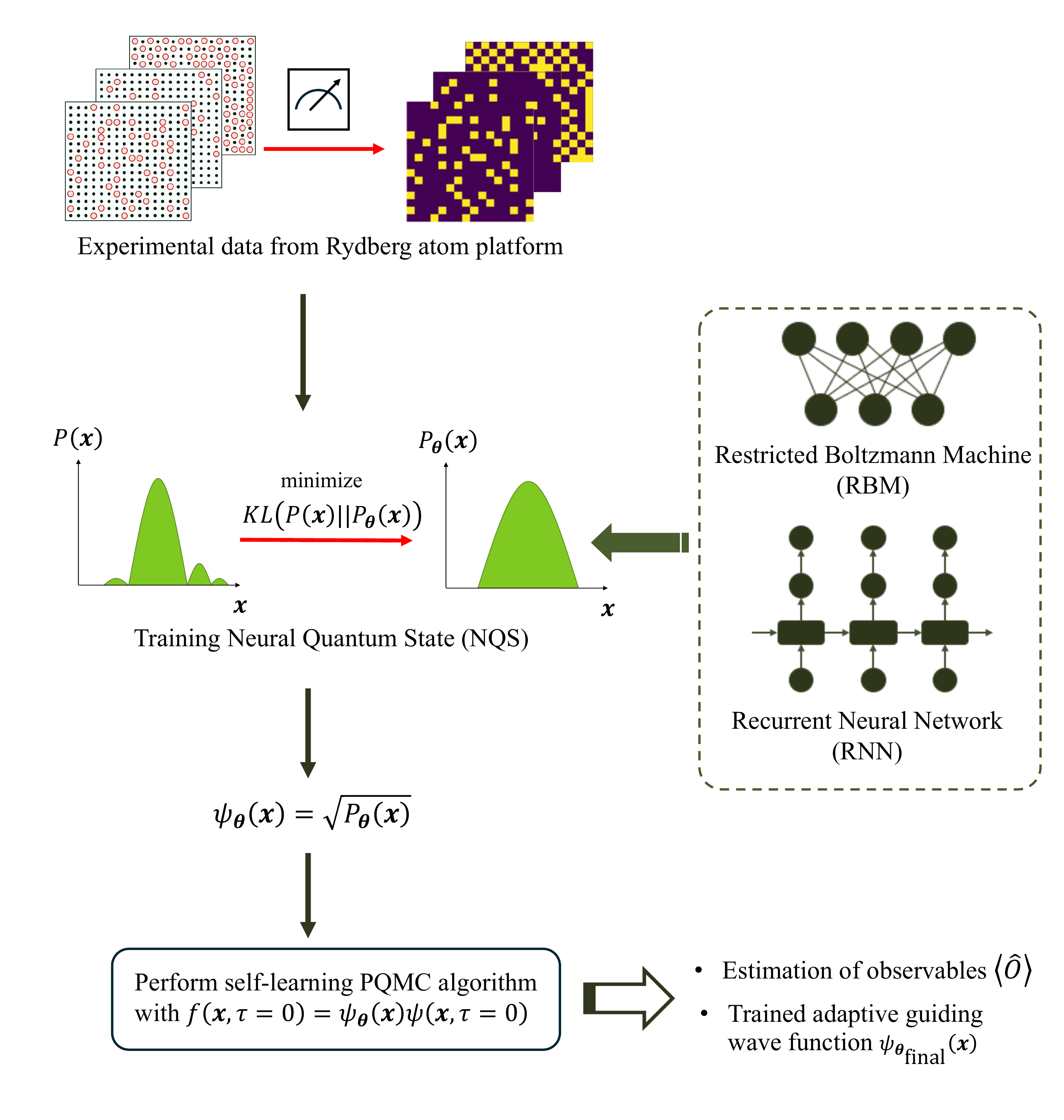}
  \caption{\label{fig:workflow} \textbf{Workflow of data-enhanced self-learning projective quantum Monte Carlo}. Experimental projective measurements provide samples from a distribution $P(\boldsymbol{x})$, which are used to pretrain an NQS with probability distribution $P_{\boldsymbol{\theta}}(\boldsymbol{x})$ by minimizing $D_{\rm KL}\bigl(P(\boldsymbol{x})\Vert P_{\boldsymbol{\theta}}(\boldsymbol{x})\bigr)$. The pretrained NQS defines the guiding wavefunction $\psi_{\boldsymbol{\theta}}(\boldsymbol{x}) =\sqrt{P_{\boldsymbol{\theta}}(\boldsymbol{x})}$ used to initialize self-learning PQMC, which iteratively refines it during the simulations. The converged simulation yields estimates of ground-state observables $\langle\hat{O}\rangle$ and a refined guiding wavefunction $\psi_{\boldsymbol{\theta}}^{\rm final}(\boldsymbol{x})$. The details of the algorithm procedure are summarized in Algorithm~\ref{alg:data_slpqmc}. }
\end{figure}

\begin{figure}[t]
\renewcommand{\figurename}{Algorithm}
\renewcommand{\thefigure}{1}
\renewcommand{\theHfigure}{alg.1}
\setlength{\abovecaptionskip}{0pt}
\setlength{\belowcaptionskip}{0pt}

\hrule height 0.8pt depth 0pt \kern 2pt
\caption{Data-enhanced self-learning PQMC}
\label{alg:data_slpqmc}
\addtocounter{figure}{-1}
\kern 2pt \hrule \kern 2pt

\begin{algorithmic}[0]
\State \textit{Setup:}
\State Initialize a probabilistic model $P_{\boldsymbol{\theta}}(\boldsymbol{x})$
\State Set the target walker population $N_w$
\State Set the number of projection iterations $N_{\rm SL}$
\State Set the number of training steps $N_{\rm train}$
\State Set the self-learning training fraction $\alpha\in(0,1)$

\State \hrulefill
\State \textit{Data pretraining:}
\State Train $P_{\boldsymbol{\theta}}(\boldsymbol{x})$ on experimental samples $\boldsymbol{x}\sim P(\boldsymbol{x})$ by minimizing $D_{\rm KL}\bigl(P(\boldsymbol{x})\Vert P_{\boldsymbol{\theta}}(\boldsymbol{x})\bigr)$ (see Fig.~\ref{fig:workflow})
\State \hrulefill
\State \textit{Self-learning PQMC:}
\State Set up the NQS guiding wavefunction
$\psi_{\boldsymbol{\theta}}(\boldsymbol{x})=\sqrt{P_{\boldsymbol{\theta}}(\boldsymbol{x})}$

\For{$n=1,\dots,N_{\rm SL}$}
    \State Propagate $N_w$ walkers with PQMC using Eq.~(\ref{eq:is_pqmc_update_rule})
    \If{$n\le\alpha N_{\rm SL}$}
        \State Build a training set of walker configurations $\mathcal{D}_{\rm PQMC}$
        \State Update $\boldsymbol{\theta}$ by minimizing
        $D_{\rm KL}(P_{\rm PQMC}\Vert P_{\boldsymbol{\theta}})$ over
        \Statex \hspace{\algorithmicindent} $N_{\rm train}$ steps
    \EndIf
\EndFor
\State \hrulefill
\end{algorithmic}
\end{figure}

\subsection{Data-enhanced self-learning PQMC}
\label{sec:data_slpqmc}

PQMC studies the ground state properties of the Hamiltonian $\hat{H}$ by stochastically evolving the state $\ket{\psi\left(\tau\right)}$ over the imaginary time step $\Delta\tau$ according to the update rule:

\begin{equation}
\label{eq:spqmc_update_rule}
    \psi\left(\boldsymbol{x},\tau+\Delta\tau\right)
    =\sum_{\boldsymbol{x'}}G\left(\boldsymbol{x},\boldsymbol{x'},\Delta\tau\right)\psi\left(\boldsymbol{x'},\tau\right),
\end{equation}
where $\psi\left(\boldsymbol{x},\tau\right)$ is the projection of the trial state onto the computational basis state $\ket{\boldsymbol{x}}=\ket{x_1, x_2, ..., x_N}$, and $G\left(\boldsymbol{x},\boldsymbol{x'},\Delta\tau\right) = \langle\boldsymbol{x}|
e^{-\Delta\tau(\hat{H}-E_{\mathrm{T}})}
|\boldsymbol{x}'\rangle$ is the Green's function corresponding to the operator $\hat{H}$. $E_{\mathrm{T}}$ is a term stabilizing the simulations. Provided that the initial state has nonzero overlap with the ground state, repeated application of Eq.~(\ref{eq:spqmc_update_rule}) suppresses excited-state contributions and yields the ground state wavefunction $\phi_0(\boldsymbol{x})$ in the long-imaginary-time limit \cite{Becca17,foulkes01}. In this work, we employ a continuous-time implementation, thereby avoiding a finite Trotter time-step error \cite{Becca17,sorella00}. In practice, $G\left(\boldsymbol{x},\boldsymbol{x'},\Delta\tau\right)$ is normalized to represent a properly defined stochastic matrix for the Monte Carlo simulations. This is done at the cost of having to evolve a large number of walkers $N_w$ that are reweighted by its normalization factor in a birth and death process called branching. The latter can hamper the efficiency of the PQMC simulations as the required number of walkers $N_w$ can scale exponentially with the system size~\cite{nemec10}.
The scaling $N_w$ with the system size can be greatly reduced by implementing importance sampling with a guiding wavefunction
$\psi_G(\boldsymbol{x})$ in which a mixed distribution $f(\boldsymbol{x},\tau) = \psi_G(\boldsymbol{x})\psi(\boldsymbol{x},\tau)$ is now evolved in imaginary-time via the equation:
\begin{equation}
\label{eq:is_pqmc_update_rule}
    f \left(\boldsymbol{x},\tau+\Delta\tau\right)
    =\sum_{\boldsymbol{x'}}\widetilde{G}\left(\boldsymbol{x},\boldsymbol{x'},\Delta\tau\right)f\left(\boldsymbol{x'},\tau\right),
\end{equation}
where the Green's function with importance sampling is given by $\widetilde{G}(\boldsymbol{x},\boldsymbol{x}',\Delta\tau)=G(\boldsymbol{x},\boldsymbol{x}',\Delta\tau) \frac{\psi_G(\boldsymbol{x})}{\psi_G(\boldsymbol{x}')}$. The distribution
$f$ is represented by $N_w$ walkers, whose branching accounts for
the local weights now associated with the modified Green's function. If $\psi_G\propto\phi_0$, the local energy $E_{\rm loc}(\boldsymbol{x})=\frac{\langle\boldsymbol{x}|\hat{H}|\psi_G\rangle}{\langle\boldsymbol{x}|\psi_G\rangle}$ used to approximate the ground state energy is configuration independent and branching fluctuations vanish leading to the need of a single walker to accurately recover ground state properties. For an imperfect
$\psi_G(\boldsymbol{x})$, these fluctuations lead to population-control bias at finite
$N_w$ with a more favourable scaling than the case without importance sampling~\cite{nemec10,boninsegni12}. Although this bias can be eliminated by sufficiently increasing $N_w$, improving the guiding wavefunction provides a more efficient
route toward more stable and faster converging simulations. Expectation values of the Hamiltonian and of observables $\hat{\mathcal{O}}$ commuting with it are evaluated via the mixed estimator $\langle\hat{\mathcal{O}}\rangle_{ \rm mixed} = \frac{\langle\phi_0|\hat{\mathcal{O}}|\psi_G\rangle}{\langle\phi_0|\psi_G\rangle}$.  For such operators, the mixed estimator coincides with the pure estimator, provided that population-control bias is eliminated and the imaginary-time projection has converged.

Self-learning PQMC achieves this by representing the guiding wavefunction with an
NQS such that, $\psi_G(\boldsymbol{x}) \equiv \psi_{\boldsymbol{\theta}}(\boldsymbol{x})$ with the NQS parameters $\boldsymbol{\theta}$ trained from the walkers distribution generated during the projection step~\cite{pilati19}. For the stoquastic Rydberg Hamiltonian considered here, positive-definite amplitudes can be used, such that $
\psi_{\boldsymbol{\theta}}(\boldsymbol{x})=\sqrt{P_{\boldsymbol{\theta}}(\boldsymbol{x})}$ where $P_{\boldsymbol{\theta}}$ is the probability distribution
represented by the neural network guiding wavefunction. Configurations generated by the projected walker population define an empirical distribution
$P_{\mathrm{PQMC}}$, from which the guiding wavefunction is updated
by minimizing $
D_{\mathrm{KL}}
\left(
P_{\mathrm{PQMC}}
\,\Vert\,
P_{\boldsymbol{\theta}}
\right)$, which is the KL divergence between the walkers and the NQS distribution. 
An important property of this procedure is that the dependence on
the initial guiding wavefunction is rapidly suppressed. In the
ideal limit in which each projection stage samples the stationary
mixed distribution and the NQS learns it exactly, the guiding wavefunction after
$n$ self-learning updates satisfies $
\psi_{\boldsymbol{\theta},n}(\boldsymbol{x})
\propto
\phi_0(\boldsymbol{x})^{1-2^{-n}}
\psi_{\boldsymbol{\theta},0}(\boldsymbol{x})^{2^{-n}}. $
The influence of the initial guide is therefore exponentially suppressed, with
$\psi_{\boldsymbol{\theta},n}\rightarrow\phi_0$ as $n\rightarrow\infty$.
This also motivates restricting NQS optimization to the first $\alpha N_{\mathrm{SL}}$ PQMC steps, with $\alpha\in[0,1]$. Once the NQS has sufficiently adapted to the projected distribution, continued training provides diminishing returns.

The exponential suppression of the initial guiding wavefunction does not, however, imply that its quality is irrelevant. The above argument assumes that each projection stage accurately samples the corresponding importance-sampled mixed distribution. At finite $N_w$, the quality of this sampling depends on $\psi_G$. Thus, a guiding wavefunction closer to the ground state reduces branching fluctuations and population-control bias, yielding a more faithful finite-population representation of the projected mixed distribution even before the first self-learning update. A better initial NQS can therefore improve the samples on which subsequent self-learning is built, despite its direct influence being progressively suppressed.

Here, we exploit experimental projective measurements to provide such an informed initialization by pretraining the NQS before self-learning PQMC begins. Previous work has shown that measurements from programmable Rydberg-atom simulators can improve NQS initialization and subsequent VMC optimization \cite{moss24}. We investigate whether a similar advantage carries over to PQMC, where the role of the data is fundamentally different. In data-enhanced VMC, the pretrained NQS is subsequently refined through variational energy minimization. In the \textit{data-enhanced self-learning PQMC} introduced here, it instead warm starts the NQS for importance sampling, after which Hamiltonian-driven imaginary-time projection generates progressively improved walker distributions used to retrain the NQS. Experimental pretraining can thus improve the initial importance-sampled projection, while the subsequent self-learning dynamics progressively reduces dependence on the imperfect experimental data.

The complete protocol is illustrated in Fig.~\ref{fig:workflow} and summarized in Algorithm~\ref{alg:data_slpqmc}. Once the walker population has converged, the ground-state energy is evaluated using the standard mixed estimator. For observables that do not commute with $\hat{H}$, such as the checkerboard order parameter considered below, we instead employ pure estimators obtained using the standard forward-walking technique \cite{casulleras95}. Simulation parameters are summarized in Appendix~\ref{app:pqmcparams}.
\subsection{Neural-network guiding wavefunctions}
\label{sec:nqs}

We employ two NQS architectures as guiding wavefunctions: a restricted Boltzmann machine~\cite{carleo17,melko19} and a recurrent neural network~\cite{hibat-allah20, wah2026autoregressiveprojectivequantummonte}. These architectures allow us to assess how the expressivity and resulting quality of the guiding wavefunction affect the efficiency of data-enhanced self-learning PQMC.

An RBM is an energy-based model defining a joint probability distribution $P(\boldsymbol{x},\boldsymbol{h})$ over binary visible variables $\boldsymbol{x}=\left(x_1,...,x_{N}\right)$ and hidden variables $\boldsymbol{h}=(h_1,\ldots,h_{N_h})$. The two layers are coupled through the energy function
\begin{equation}
\label{eq: rbm_hamiltonian}
    H_{\rm RBM}\left(\boldsymbol{x}, \boldsymbol{h}\right)=-\sum_{i,j}J_{i,j}h_ix_j-\sum_ja_jx_j-\sum_ib_ih_i,
\end{equation}
where the visible biases $a_j$, hidden biases $b_i$, and weights $J_{ij}$ constitute the trainable parameters $\boldsymbol{\theta}$. For the stoquastic Rydberg Hamiltonian considered here, the ground-state amplitudes can be chosen real and positive, and we therefore restrict $\boldsymbol{\theta}$ to real values. Marginalizing over the hidden variables gives
\begin{equation}
\label{eq:rbm}
P_{\rm RBM}(\boldsymbol{x})
\propto
e^{\sum_j a_jx_j}
\prod_i
2\cosh\left(
b_i+\sum_jJ_{ij}x_j
\right).
\end{equation}
The RBM guiding wavefunction is then defined as $\psi_{\boldsymbol{\theta}}(\boldsymbol{x}) =\sqrt{P_{\rm RBM}(\boldsymbol{x})}$, and its parameters are optimized using $k$-step contrastive divergence~\cite{hinton02} and stochastic gradient descent with momentum.

RNN wavefunctions are autoregressive models that represent the probability distribution associated with a positive wavefunction as a product of normalized conditional probabilities~\cite{hibat-allah20}. In Ref.~\cite{wah2026autoregressiveprojectivequantummonte}, they were used as guiding wavefunctions of one-dimensional spin chains. In this work, we use them in two dimensions. In particular, we employ the patching strategy of Ref.~\cite{sprague24}, which has been shown to improve the efficiency of RNN wavefunctions for Rydberg Hamiltonians on square lattices. For a two-dimensional configuration $\boldsymbol{x}$, the lattice is decomposed into a raster-ordered sequence of spatial patches ${\boldsymbol{x}_{\mathcal{P}i}}$, and the probability distribution is factorized as
\begin{equation}
\label{eq:rnn}
P_{\rm RNN}(\boldsymbol{x})=\prod_i p_{\boldsymbol{\theta}}\bigl(\boldsymbol{x}_{\mathcal{P}_i}\bigm|\boldsymbol{x}_{\mathcal{P}_{<i}}\bigr),
\end{equation}
where $\mathcal{P}_i$ denotes the sites belonging to the $i$-th patch and $\mathcal{P}_{<i}$ all preceding patches. Each conditional probability $p_{\boldsymbol{\theta}}\bigl(\boldsymbol{x}_{\mathcal{P}_i}\bigm|\boldsymbol{x}_{\mathcal{P}_{<i}}\bigr)$ is produced by a recurrent cell and normalized via a Softmax function, making $P_{\rm RNN}(\boldsymbol{x})$ normalized by construction. We use $2\times2$ patches throughout this work and train the RNNs using batch gradient descent with the Adam optimizer \cite{kingma2014adam}.

Although both the RBM and RNN define generative models, sampling of physical spin configurations from them is not required within self-learning PQMC. Instead, they are used as function approximators to evaluate the guiding wavefunction during the projective step and the NQS probabilities required for training, while configurations are generated by the PQMC walker dynamics. Nevertheless, RNNs provide an additional advantage given that, once trained, they can generate exact samples via ancestral sampling. Details of the network architectures, optimization, and hyperparameters are provided in Appendix~\ref{app:nqsparams}.

\section{Results}
\label{sec:results}

We now evaluate the performance of data-enhanced self-learning PQMC for the
two-dimensional Rydberg array introduced in Sec.~\ref{sec:rydberg}. We first investigate whether pretraining an NQS with experimental data improves the efficiency of self-learning PQMC by reducing finite-walker effects and accelerating the convergence of the self-learning procedure. We then examine the accuracy of the method across the disordered-to-checkerboard transition, considering both the ground-state energy and the checkerboard order parameter. Finally, we independently benchmark self-learning PQMC through a finite-size scaling analysis of the phase transition and the extraction of its critical properties. We benchmark our results against another quantum Monte Carlo technique implemented with stochastic series expansion (SSE)~\cite{merali24} which is considered to provide exact ground state properties within statistical accuracy.

\subsection{\label{sec:level2a} Improving self-learning PQMC efficiency with experimental data}
\begin{figure}
  \centering
  \includegraphics[width=\columnwidth]{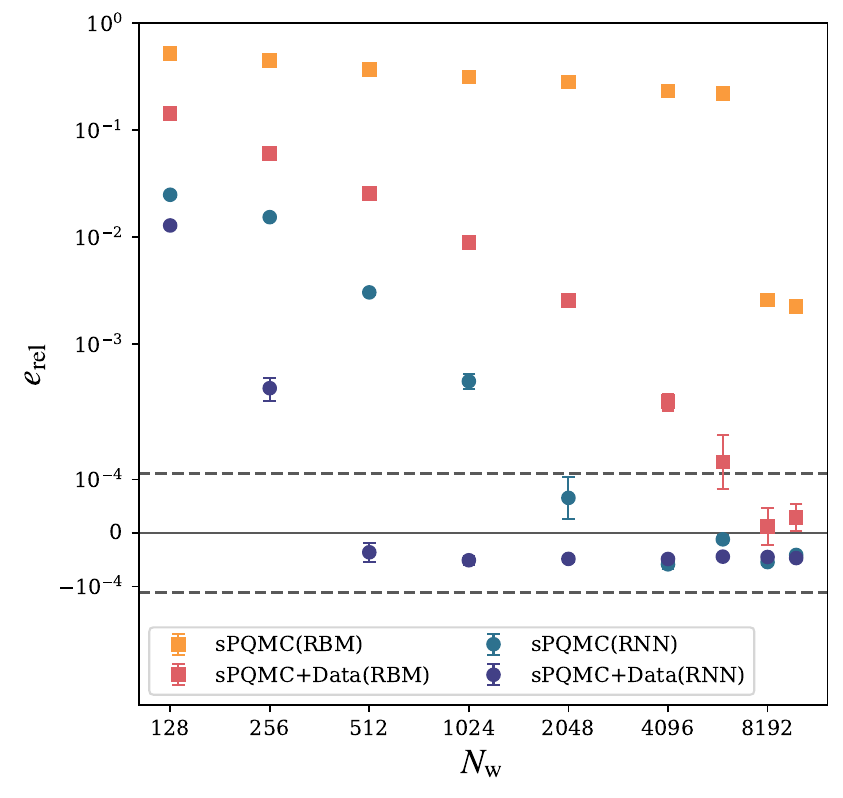}
  \caption{\label{fig:energy_vs_walker} {Relative ground-state energy versus walker population. Signed relative energy $e_{\rm rel}$ between the ground state energies obtained from self-learning PQMC (sPQMC) and data-enhanced self-learning PQMC (sPQMC+Data) using RBM and RNN guiding wave functions, compared to the SSE reference energy, as a function of the walker population $N_{\rm w}$. The simulations are performed on a $16\times16$ lattice close to the phase transition at a detuning $\Delta/\Omega=1.051$. The horizontal dashed lines indicate the positive and negative bounds of the SSE statistical uncertainty.}}
\end{figure}

\begin{figure}
  \centering
  \includegraphics[width=\columnwidth]{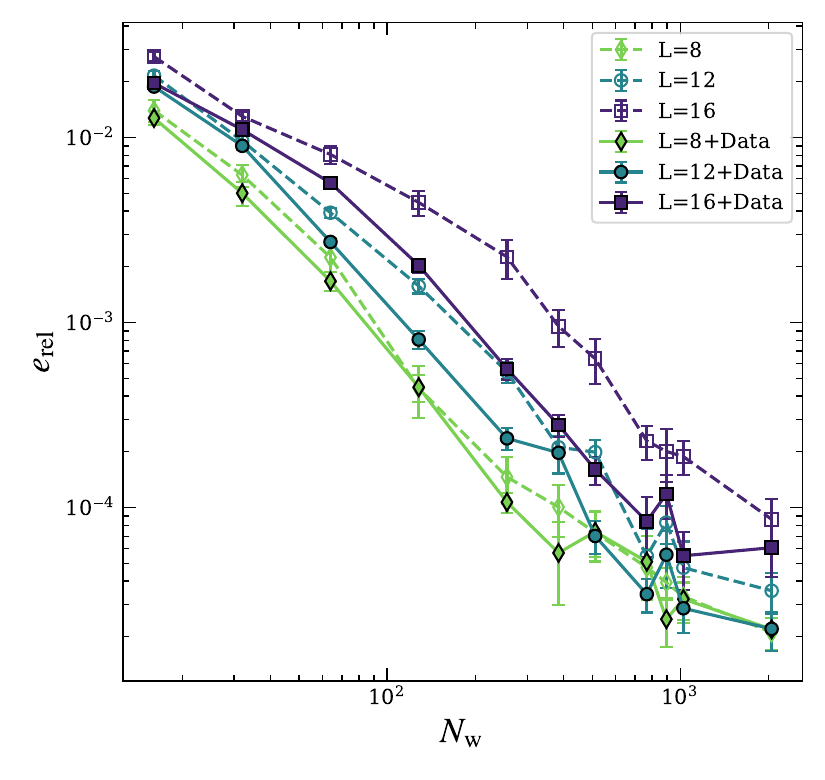}
  \caption{\label{fig:system_size_scalings}\textbf{Walker population dependence on the system size.}
Energy relative error $e_{\rm rel}$ as a function of $N_{\rm w}$ at different system sizes of the $L \times L$ Rydberg array. Empty symbols represent the sPQMC results while filled symbols represent sPQMC+Data results using the RNN guiding wavefunction. The reference energy used to compute $e_{\rm rel}$ is the mixed estimator energy of the corresponding method at $N_{\rm w} = 10000$. Simulations are performed at $\Delta/\Omega=1.051$.}
\end{figure}

\begin{figure}
  \centering
  \includegraphics[width=\columnwidth]{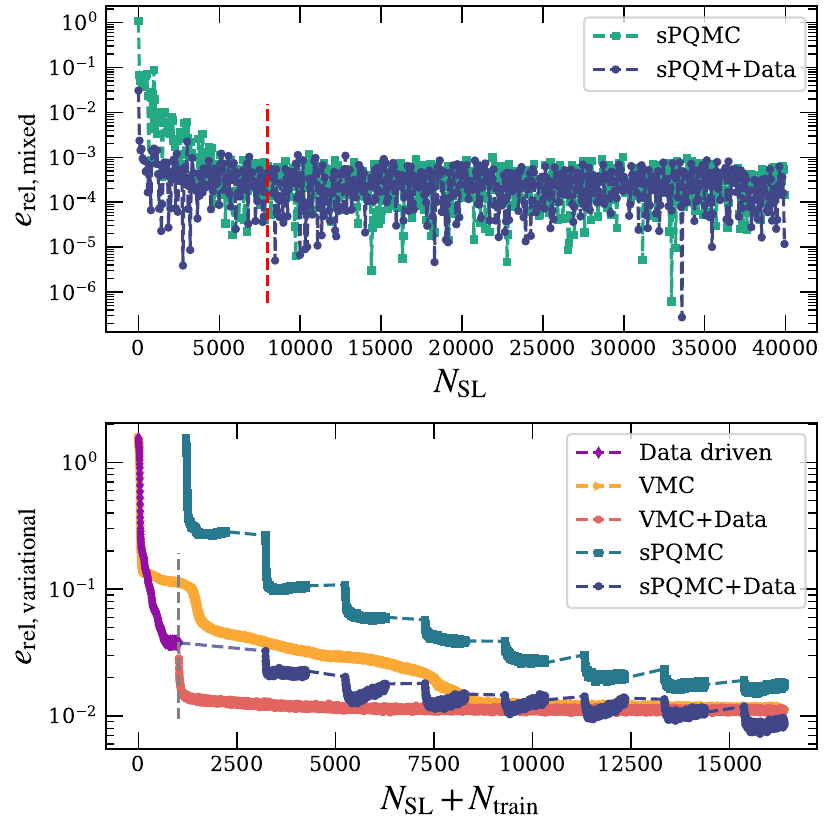}
  \caption{\label{fig:relative_error_convergence}\textbf{Convergence of the ground-state energy.}
The top panel shows the energy relative error of the PQMC mixed estimator
as a function of the self-learning iteration $N_{\rm SL}$ for sPQMC
and sPQMC+Data. The bottom panel shows the variational energy error
of the corresponding NQS and of NQSs trained with variational Monte
Carlo (VMC). Data driven denotes the NQS pretrained on
experimental data, while VMC+Data denotes VMC initialized from the
data-pretrained NQS. VMC is trained for the
same cumulative number of training steps, $N_{\rm SL}+N_{\rm train}$. The vertical dashed line indicates the transition from Data-Driven to both VMC+Data and sPQMC+Data. The dashed horizontal lines in the PQMC results denote the portion of the simulation where the PQMC simulations are carried out with a fixed guiding wavefunction. 
Only the first $20\%$ of the simulations, during which the NQS
is trained, are shown (left of the red vertical dashed line). Simulations are performed on a
$16\times16$ lattice at a detuning $\Delta/\Omega=1.051$, using RNN guiding wavefunctions and $N_w=4096$ walkers.
}
\end{figure}

\begin{figure}
  \centering
  \includegraphics[width=\columnwidth]{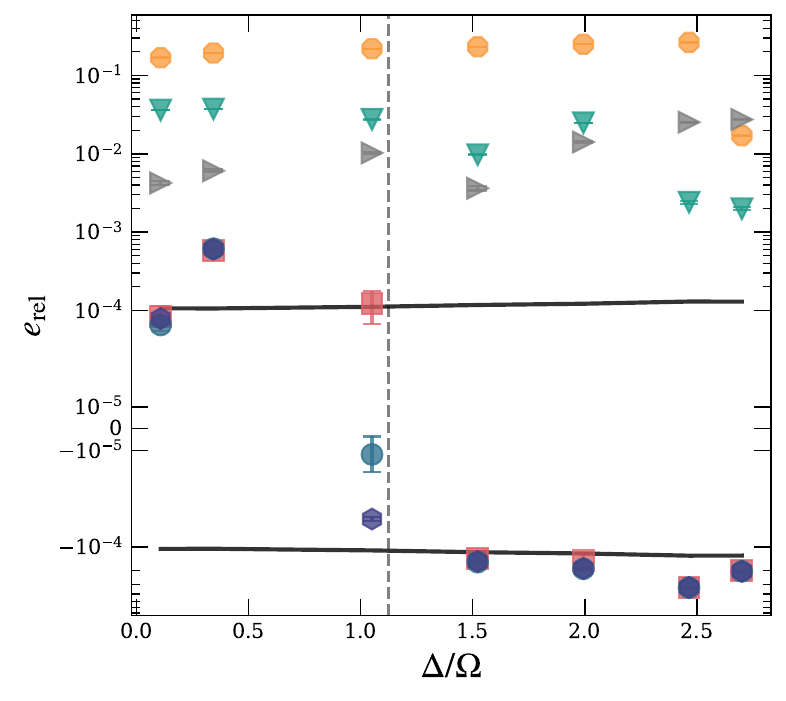}
  \caption{\label{fig:relative_error_phase_diagram}\textbf{Relative ground-state energy across the phase diagram.}
Signed relative energy difference $e_{\rm rel}$
between the ground-state energies obtained from the different methods and
the SSE reference energy, as a function of
the detuning $\Delta/\Omega$. Results are shown for sPQMC and sPQMC+Data
using RBM and RNN guiding wavefunctions, and compared with VMC on a
$16\times16$ lattice. PQMC simulations are performed with $N_w=6000$
walkers. The vertical dashed line indicates the critical detuning
$\Delta_c/\Omega\approx1.12(4)$~\cite{ebadi21}.}
\end{figure}

We first investigate how experimental pretraining affects the walker population required to obtain unbiased estimation of ground state energies. We quantify the accuracy of the simulations using relative energy, which is defined as:
\begin{equation}
    \label{eq:relative_error}
    e_{\rm rel}
    = \left| \frac{\langle\hat{H}\rangle - E_{\rm ref}}{E_{\rm ref}} \right|,
\end{equation}
where $E_{\rm ref}$ is a reference energy.
Figure~\ref{fig:energy_vs_walker} shows the relative energy difference
defined as in Eq.~\eqref{eq:relative_error}, but retaining the sign of the numerator and the modulus in the denominator
as a function of the number of walkers $N_w$ for self-learning PQMC (sPQMC) with RBM and
RNN guiding wavefunctions, with and without experimental-data pretraining.
The simulations are performed on the $16\times16$ array close to the phase
transition, at $\Delta/\Omega=1.051$, where an accurate guiding wavefunction
is particularly important. The sign indicates whether the estimated energy lies above or below the SSE (the absolute value is kept in the denominator). In all cases, increasing $N_w$ systematically
reduces the finite-population bias toward the SSE reference energy. Experimental pretraining accelerates
this convergence, allowing accurate energies to be obtained with fewer
walkers.
The improvement is particularly pronounced for the RBM guiding wavefunction. Without pretraining, RBM-guided sPQMC does not reach 
the accuracy within the uncertainty of the SSE reference within the range of $N_w$ considered in Fig.~\ref{fig:energy_vs_walker},
whereas its data-enhanced variant, sPQMC+Data, approaches the reference with
substantially fewer walkers. Importantly, this does not reflect an intrinsic
limitation of RBM-guided PQMC. The finite-population bias that causes the discrepancy
can, in principle, be systematically reduced by increasing $N_w$, as demonstrated in
Fig.~\ref{fig:walker_convergence} of
Appendix~\ref{app:walker_convergence}, albeit at increased computational cost.
Improving the guiding wavefunction, therefore, provides a more efficient route
to suppressing finite-population effects. The RNN provides a considerably
more accurate guiding wavefunction even without experimental pretraining, reaching the SSE
result with fewer walkers than the data-pretrained RBM. Experimental
pretraining, nevertheless, further reduces the walker population required by
the RNN-guided simulations. These results show that the efficiency gain from
experimental initialization depends on the quality of the underlying guiding
wavefunction.

We further examine the impact of a finite walker population on the RNN guiding wave function by studying its effect on different lattice sizes. In Fig.~\ref{fig:system_size_scalings}, $E_{\rm ref}$ is taken to be the average energy of the corresponding method at $N_w=10000$ walkers. In order to obtain training datasets for the $L=8$ and $L=12$ system sizes (given that we only had access to experimental data on a $16 \times 16$ lattice), we cropped projective measurement configurations from the $L=16$ experimental array while preserving the relative populations of atoms in the Rydberg state $\ket{r}$ and the ground state $\ket{g}$. We observe that, as the system size increases, the number of walkers required to keep increasingly lower relative errors grows faster for sPQMC compared to sPQMC+Data, eventually plateauing in the larger number of walker limit. These results suggest that the walker population required to achieve a given accuracy grows more rapidly with system size without experimental pretraining, indicating that the relative benefit of sPQMC+Data may become increasingly important for larger systems.


We next investigate how experimental pretraining affects the convergence of
the self-learning procedure with respect to the number PQMC steps. For the remainder of the manuscript, we quantify the accuracy using the relative
energy error with $E_{\rm ref}$ being the corresponding SSE energy.
The top panel of Fig.~\ref{fig:relative_error_convergence} shows the relative
error of the PQMC mixed estimator $\langle\hat{H}\rangle_{ \rm mixed}$ 
as a function of the number of self-learning iterations $N_{\rm SL}$.

Here, we use RNN guiding wavefunctions with $N_w=4096$.
Data-enhanced sPQMC starts from a smaller energy error and converges more
rapidly toward the SSE reference than sPQMC initialized without experimental
data. The advantage is most pronounced during the early stages of the
simulation and progressively decreases as the guiding wavefunctions are
refined using the projected walker distributions. This behavior is consistent
with the self-learning mechanism discussed in Sec.~\ref{sec:data_slpqmc}:
although the influence of the initial guiding wavefunction is progressively
suppressed, a better initialization improves the quality of the projection
during the early self-learning iterations.

The bottom panel of Fig.~\ref{fig:relative_error_convergence} instead tracks
the variational energy of the corresponding RNN guiding wavefunctions,
$\langle\hat{H}\rangle_{{\boldsymbol{\theta}}}=
\langle\psi_{\boldsymbol{\theta}}|\hat{H}|
\psi_{\boldsymbol{\theta}}\rangle/
\langle\psi_{\boldsymbol{\theta}}|\psi_{\boldsymbol{\theta}}\rangle$,
and compares it with RNNs with the same architecture optimized directly using variational Monte Carlo (VMC) with Adam optimizer. For the projective
simulations, samples from the RNN are generated only to evaluate the
variational energy; the guiding wavefunction itself remains trained by
unsupervised learning from the PQMC walker distribution. Experimental
pretraining already provides an NQS with a substantially smaller variational
energy error than a randomly initialized model. Interestingly, after the 
following the second self-learning
update, over the interval $N_{\rm SL}+N_{\rm train}\approx 5000$ to
$12500$, the variational energy temporarily rises while
$D_{\mathrm{KL}}(P_{\mathrm{PQMC}}\Vert P_{\boldsymbol{\theta}})$
decreases (not shown). This is not contradictory, since the self-learning optimization
does not minimize the variational energy directly, but instead trains the NQS
to reproduce the projected walker distribution. Subsequent self-learning
updates lower the variational energy again, driving the randomly initialized NQS
towards the accuracy obtained with data-enhanced VMC, while the pre-trained NQS achieves a smaller 
variational energy error than data-enhanced VMC.
It is also worth noting that evaluating the Kullback–Leibler divergence loss for patched RNN training has a computational complexity of $\mathcal{O}(N)$, compared with $\mathcal{O}(N^2)$, up to the prefactor patch size, for the local-energy calculations required for PQMC projection and VMC optimization, making the KL-based training objective computationally less expensive. We report the execution times of the data-enhanced sPQMC in Appendix~\ref{app:exec_time}.

The behavior of sPQMC initialized from a random NQS provides a complementary
picture. Its guiding wavefunction initially has a substantially larger
variational energy error, but improves rapidly under successive self-learning
updates and eventually outperforms the corresponding VMC optimization,
approaching the accuracy of the data-enhanced methods. This further supports
the role of experimental data primarily as an informative initialization:
it provides a better initial guiding wavefunction and hence a more accurate
finite-walker representation of the projected distribution, accelerating the
subsequent self-learning dynamics. Finally, Fig.~\ref{fig:relative_error_convergence} also illustrates the
distinction between the accuracy of the guiding wavefunction and that of the
projective calculation itself. Even when the guiding NQS remains imperfect,
the PQMC mixed estimator can provide substantially more accurate ground-state
energies than the variational energy of the same NQS, because imaginary-time
projection suppresses its remaining excited-state components.

We next assess whether the improvement observed at $\Delta/\Omega=1.051$
persists across the disordered-to-checkerboard transition.
Figure~\ref{fig:relative_error_phase_diagram} shows the signed relative energy difference
for RBM- and RNN-guided sPQMC with $N_w=6000$ walkers.
For comparison, we also report the corresponding VMC results and the
statistical uncertainty of the SSE estimates. RBM neural quantum states implemented within the VMC simulations are optimized via stochastic reconfiguration, while, for RNNs, we use the Adam optimizer.
For the RBM guiding wavefunctions, experimental pretraining produces a substantial improvement
over the entire range of detunings considered, reducing the magnitude of the
energy error by more than one order of magnitude in most cases. The self-learning procedure in combination with data pretraining proves to be the best recipe for optimal results in this setting as demonstrated by ablation studies in Appendix~\ref{app:fixed_guide_comparison} that isolate their respective effects. 
The improvement
is considerably smaller for the RNN, for which sPQMC already achieves high
accuracy without experimental pretraining. Nevertheless, data-enhanced
RNN-guided sPQMC remains among the most accurate calculations across the
transition. Note that though some of the PQMC estimates lie slightly below the SSE reference
energy, they are still within SSE errorbars.
\subsection{\label{sec:level2b} Study of the disordered-to-checkerboard phase transition}

We now turn to the disordered-to-checkerboard phase transition. We
characterize it through the Fourier component of the local density at $(k_x,k_y)=(\pi,\pi)$,
\begin{equation}
    \label{eq:fourier_transform}
    \mathcal{F}_{\pi,\pi}
    =\frac{1}{N}\sum_i \hat{n}_i
    \exp\left[i(\pi,\pi)\cdot(x_i,y_i)\right].
\end{equation}
Unlike the energy, this observable does not commute with the Hamiltonian and
therefore cannot be evaluated without bias directly from the mixed
distribution. We instead employ forward walking, propagating the walkers through additional imaginary-time steps before each measurement to suppress the dependence on the guiding wave function. In this work, we find that an imaginary-time propagation interval of $200$ is sufficient for our calculations.

\begin{figure}
  \centering
  \includegraphics[width=\columnwidth]{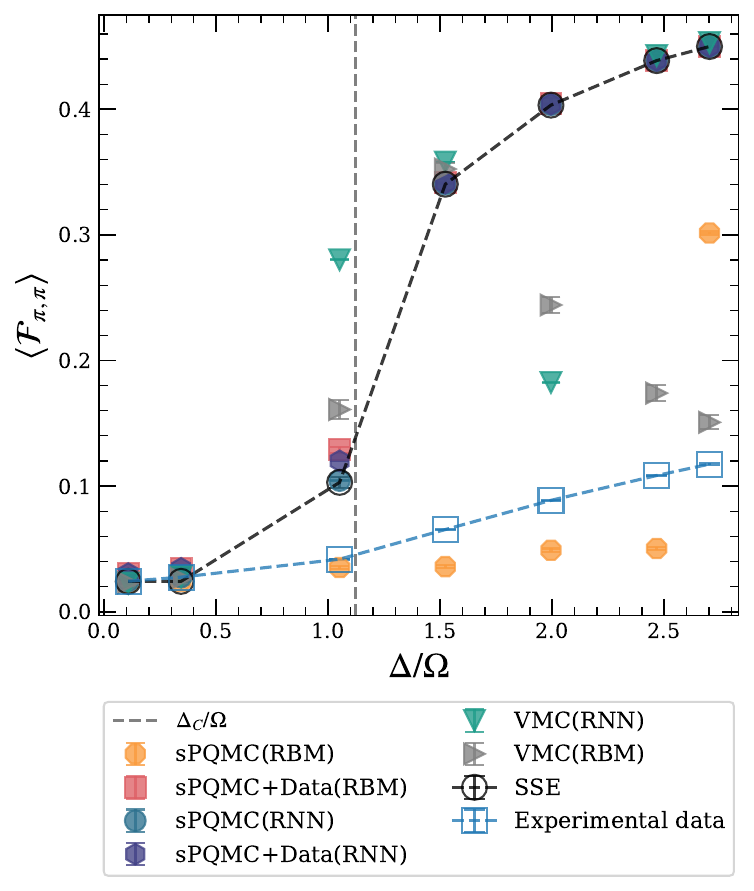}
  \caption{\label{fig:phase_diagram}\textbf{Checkerboard order parameter across the phase transition.}
  Checkerboard order parameter $\mathcal{F}_{\pi,\pi}$ as a function of the
  detuning $\Delta/\Omega$. The blue squares represent experimental data from
  Ref.~\cite{ebadi21}, while the black circles show the SSE reference results.
  The remaining data are obtained using the same simulation protocol as in
  Fig.~\ref{fig:relative_error_phase_diagram}. The dotted lines are guides to
  the eye.}
\end{figure}

Figure~\ref{fig:phase_diagram} shows the resulting checkerboard order
parameter computed as $\langle |\mathcal{F}_{\pi,\pi}| \rangle$ across the disordered-to-checkerboard transition. The simulations
follow the same protocol as in Fig.~\ref{fig:relative_error_phase_diagram}
and are compared with SSE and the experimental measurements. The benefit of
experimental pretraining observed for the ground-state energy persists for
the order parameter. In particular, the improvement is substantial for the
RBM-guided simulations, which otherwise show appreciable deviations from the
SSE results near the transition and within the checkerboard phase. The
data-enhanced RBM results are instead brought considerably closer to the SSE
benchmark. The RNN-guided simulations are already more accurate without
pretraining, while the inclusion of experimental data provides a further,
although smaller, improvement. This behavior is consistent with the
architecture dependence observed for the energy in
Figs.~\ref{fig:energy_vs_walker} and
\ref{fig:relative_error_phase_diagram}.

Interestingly, the experimental measurements themselves show visible
deviations from the equilibrium numerical results, particularly in the
ordered phase, consistent with defects generated during the finite-rate
experimental state preparation~\cite{ebadi21}. Nevertheless, using these
measurements to initialize the guiding wavefunction improves the subsequent
sPQMC calculation, as it was also the case for data-enhanced VMC~\cite{moss24}. This illustrates an important feature of the
data-enhanced approach: the experimental samples need not provide a very accurate
representation of the ground state themselves to constitute a useful
initialization. The subsequent imaginary-time projection can refine the
initial distribution toward the ground state. We also show in Appendix~\ref{app:offdiag} that this feature remains true in the more challenging case of accessing off-diagonal observables.

Having established the benefit of experimental initialization, we finally
assess whether sPQMC can independently reproduce the critical behavior of the
model. For this purpose, we perform a finite-size scaling analysis using
sPQMC without experimental-data pretraining, thereby separating the accuracy
of the underlying projective method from the data enhancement discussed
above. We consider square lattices with $L=6,8,10,12,$ and $14$ and determine
the critical detuning from the checkerboard order parameter. 
The order parameter is expected to scale by the system size $L$ by the relation

\begin{equation}
    \label{eq:fss}
    \mathcal{F}_{\pi,\pi}
    =L^{-\beta/\nu}
    f\left[
    \frac{\Delta-\Delta_c}{\Omega}L^{1/\nu}
    \right].
\end{equation}
Here, we approximate the scaling function with a polynomial
$f\left( x \right) = f_0+f_1x+f_2x^2+...+f_nx^n$ where $f_i$ is the fitting parameter for the term $x^i$. 
We use the bootstrap method to quantify error bars on the fitted parameters. Details of fitting including polynomial order and scaling windows are reported in Appendix~\ref{app:scaling_collapse}.
The inset of Fig.~\ref{fig:fss} shows the corresponding optimal data collapse,
yielding $\Delta_c/\Omega\simeq1.148(2)$, $\beta\simeq0.330(5)$ and $\nu\simeq0.614(3)$, which is also in agreement with the critical exponents obtained from the Binder cumulants (see Appendix~\ref{app:ffs_binder}). 
Table~\ref{tab:critical_exponents} compares these estimates with previous numerical and experimental results,
as well as with the corresponding RBM-guided sPQMC simulations with $N_w=6000$. 
The RNN guiding wavefunction results are in close agreement with the DMRG estimates, while the RBM yields a slightly lower critical detuning but is consistent within statistical uncertainty; the critical exponents obtained with both architectures remain consistent with the (2+1)-dimensional Ising universality class~\cite{samajdar20}. For $R_b/a = 1.15$, we note a small discrepancy in the critical detuning between experimental results and sPQMC (RNN). This is potentially due to differences between the simulations of an idealized model and its experimental realization in an actual quantum device. We note that our choice of $N_w$ is sufficient to suppress the population control bias for the $16 \times 16$ system, as shown in  Fig.~\ref{fig:energy_vs_walker}. We therefore consider this population size to be sufficiently large to represent the large-walker-population limit for the smaller systems investigated in this analysis.

\begin{table}[t]
\scriptsize
\centering
\caption{\label{tab:critical_exponents}
Critical detuning and critical exponents of the
disordered-to-checkerboard transition. }
\begin{tabular}{lccccc}
\hline\hline
Method & $R_b/a$ & $\Delta_c/\Omega$ & $\beta$ & $\nu$ & Reference \\
\hline
PIMC
& 1.20 & $1.0959(1)$ & $0.291(1)$ & $0.632(5)$
& Ref.~\cite{kalinovski22} \\
DMRG
& 1.20 & $1.1477(6)$ & $\approx0.326$ & $\approx0.629$
& Ref.~\cite{samajdar20} \\
Experiment
& 1.15 & $1.12(4)$ & -- & $0.62(4)$
& Ref.~\cite{ebadi21} \\
\hline
sPQMC (RBM)
& 1.20 & $1.12(2)$ & $0.332(4)$ & $0.648(6)$
& \textbf{This work} \\
sPQMC (RNN)
& 1.20 & $1.148(2)$ & $0.330(5)$ & $0.614(3)$
& \textbf{This work} \\
sPQMC (RNN)
& 1.15 & $1.029(1)$ & $0.324(2)$ & $0.632(7)$
& \textbf{This work} \\
\hline\hline
\end{tabular}
\end{table}

\begin{figure}
  \centering
  \includegraphics[width=\columnwidth]{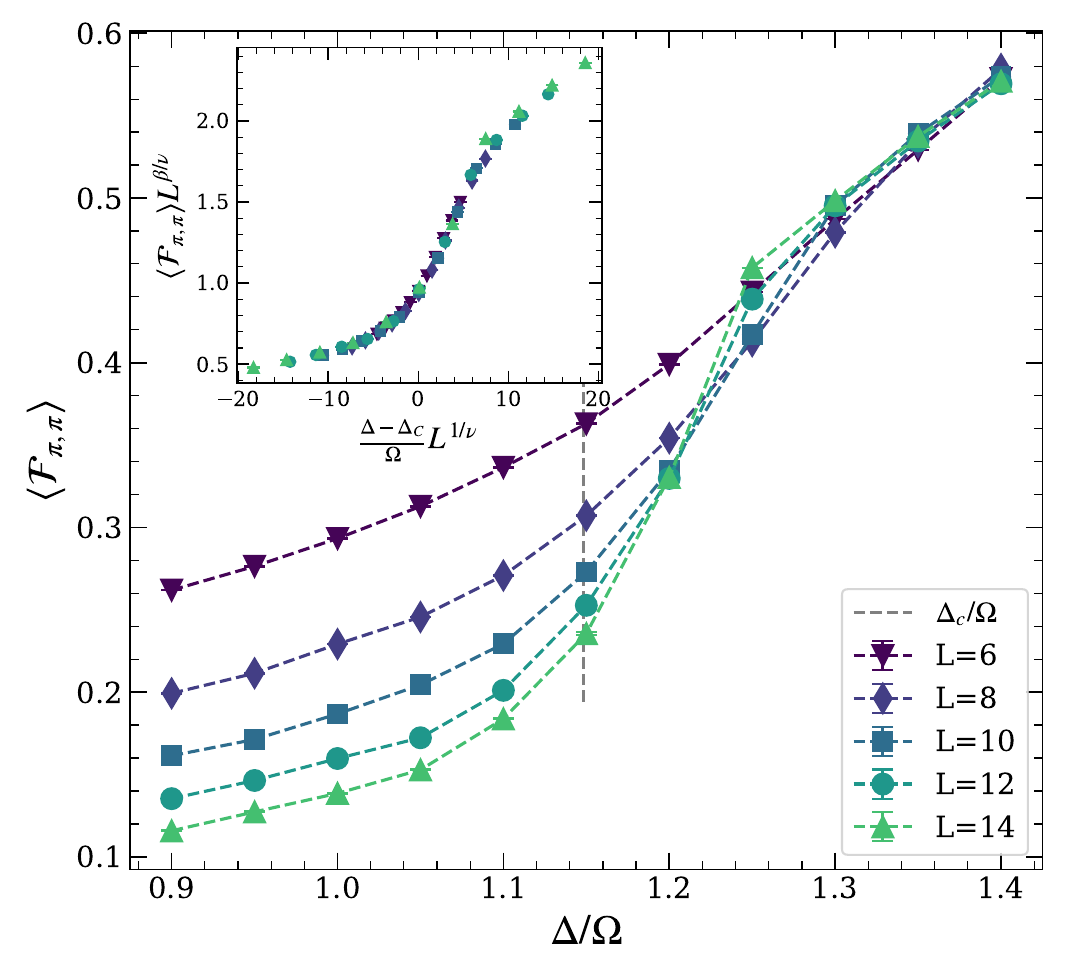}
  \caption{\label{fig:fss}\textbf{Finite-size scaling of the checkerboard phase transition.}
  Disordered-to-checkerboard phase transition order parameter $\mathcal{F}_{\pi,\pi}$ as a function of the
  detuning $\Delta/\Omega$ for different lattice sizes $L\times L$, obtained
  from sPQMC simulations without experimental-data pretraining. The inset shows the corresponding data
  collapse of the rescaled checkerboard order parameter used to extract the
  critical exponents $\nu$ and $\beta$. Finite-size scaling collapse
  yields a critical detuning
  $\Delta_c/\Omega\simeq1.147(9)$. Simulations are performed using RNN
  guiding wavefunctions with $R_b/a=1.2$ and $N_w=4096$ walkers.}
\end{figure}

\section{\label{sec:level3}Conclusion}

In this work, we introduced data-enhanced self-learning projective quantum
Monte Carlo, in which projective measurements from a programmable quantum
simulator are used to pretrain neural-network guiding wavefunctions before
the self-learning projection. We demonstrated the approach using experimental
measurements from a $16\times16$ Rydberg-atom array~\cite{ebadi21}, with both
RBM and RNN guiding wavefunctions. We have found that experimental pretraining reduces
finite-population effects and accelerates the convergence of the
self-learning procedure, allowing accurate ground-state properties to be
obtained with fewer walkers and fewer self-learning iterations. The magnitude of this improvement depends strongly on the quality of the
guiding wavefunction. For the RBM, experimental pretraining substantially
improves both the ground-state energy and the checkerboard order parameter
across the disordered-to-checkerboard transition. The RNN provides a more
accurate guiding wavefunction already without experimental initialization
and consequently exhibits a smaller, although still observable, improvement. Simulations across different system sizes further suggest that the relative benefit of experimental pretraining may increase with system size.
This architecture dependence illustrates the complementary roles of
experimental information and NQS expressivity where data can provide a
particularly useful initialization when the guiding ansatz or its
self-learning optimization alone does not yet provide an accurate
representation of the ground state for a finite walker population budget.

A notable feature of the approach is that the experimental measurements need
not themselves very accurately represent the equilibrium ground state. Indeed,
the measured checkerboard order parameter exhibits deviations from the
equilibrium numerical results, consistent with the finite-rate state
preparation of the experiment~\cite{ebadi21}. Nevertheless, these
configurations provide useful information for initializing the guiding wave
function, as previously observed in data-enhanced VMC~\cite{moss24}. In the
present approach, the subsequent imaginary-time projection further refines
this data-driven initialization where the projected walker distributions are used
to iteratively update the guiding wavefunction while the dependence on the
initial experimental distribution is progressively reduced. This interplay
between data-driven initialization and ground-state projection is the central
feature of data-enhanced self-learning PQMC. We further benchmarked self-learning PQMC independently of experimental
pretraining by extracting the critical properties of the
disordered-to-checkerboard transition. Finite-size scaling with both neural network guiding
wavefunctions yields a critical detuning and critical exponents in close
agreement with DMRG calculations~\cite{samajdar20}
and experimental data~\cite{ebadi21}.
These results demonstrate that self-learning PQMC can accurately reproduce
both ground-state observables and the critical behavior of large
two-dimensional Rydberg arrays. 

Our results motivate extending data-enhanced self-learning PQMC to more
challenging quantum many-body systems accessible to programmable quantum
simulators. Natural directions include frustrated Rydberg systems such as
Kagome lattices~\cite{samajdar21}, amorphous Rydberg arrays where glassy
physics can emerge~\cite{brodoloni25}, and quantum simulators based on trapped
ions~\cite{blatt12,smith16}. Such systems may provide particularly useful
settings for the present approach, since complex energy landscapes and
correlations can make the construction and optimization of accurate guiding
wavefunctions increasingly demanding. There is also considerable scope for improving the neural representation
used within the self-learning procedure. More expressive and scalable
architectures, including patched transformers~\cite{sprague24}, hierarchical networks~\cite{oyedemi2026scalableneuralwavefunctions},
state-space models~\cite{gu22,gu24}, and parallel-scan recurrent neural
networks~\cite{feng24, meraliarxiv26}, could further improve the quality of the
guiding wavefunction and reduce the walker population required for accurate
projection. Another possible extension is to make the framework amenable to complex guiding wavefunctions. The latter could be pretrained on experimental data as in Refs~\cite{torlai2018neural, iouchtchenko2023neural} and then employed in a self-learning setting with the fixed-node or fixed-phase approximations~\cite{ren23,  li2026scalinguniversalferminetwork}. 


\vspace{1em}


\section*{Data and code availability}
The data that support the findings of this study are available upon request from the corresponding author. 
The RNN used in the simulation is based on PyTorch \cite{paszke2019}, the code for RBM and PQMC is based on Numpy \cite{harris20}. The VMC optimization of RBM wavefunctions is performed using NetKet library \cite{carleoarxive}. We perform finite-size scaling with the code based on Scipy \cite{virtanen20}, and the visualization is performed using Matplotlib  \cite{hunter2007}.

\section{\label{sec:acknowledgment}ACKNOWLEDGEMENT}
We thank S. Czischek, M. Kohandel, and M. Hibat-Allah for their valuable discussions and insights. We acknowledge E. Merali and M. S. Moss for their invaluable support with the SSE simulations. In particular, we thank M. S. Moss for sharing the experimental data used in this work. L. B. acknowledges the support of the Perimeter Institute during his research visit. This research was supported in part by grant NSF PHY-2309135 to the Kavli Institute for Theoretical Physics (KITP).
E.M.I. acknowledges support from the Natural Sciences and Engineering Research Council of Canada (NSERC), grant no. RGPIN-2024-04407. Research at the Perimeter Institute is supported in part by the Government of Canada through the Department of Innovation, Science and Economic Development Canada and by the Province of Ontario through the Ministry of Economic Development, Job Creation, and Trade. 


\bibliography{apssamp}
\appendix
\raggedbottom

\setlength{\textfloatsep}{10pt}
\setlength{\intextsep}{10pt}
\setlength{\floatsep}{10pt}

\section{Additional benchmarks}
\label{app:additional_benchmarks}

\subsection{PQMC with fixed NQS guiding wavefunctions}
\label{app:fixed_guide_comparison}

We further benchmark data-enhanced self-learning PQMC against
alternative strategies for constructing the guiding wavefunction.
To isolate the roles of experimental initialization and self-learning,
we consider four protocols: VMC training a NQS via stochastic
reconfiguration; PQMC using the resulting variationally
optimized NQS as a fixed guiding wave functions (PQMC+VMC); PQMC using an experimentally
pretrained NQS as a fixed guiding wave functions (PQMC+Data); and data-enhanced
self-learning PQMC (sPQMC+Data), in which the experimentally pretrained
NQS is subsequently refined using configurations generated during the
projective simulation. In this subsection, we use RBM guiding wave functions.

\begin{figure}
    \centering
    \includegraphics[width=\columnwidth]{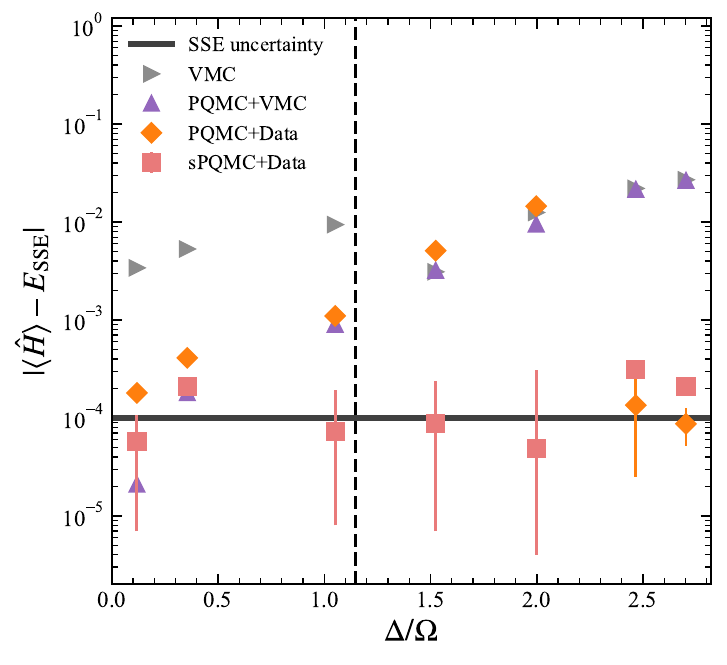}
    \caption{\label{fig:additional_energy}
    \textbf{Ground-state energy for different guiding-wave-function
    optimization strategies.}
    Ground-state energy absolute error across the phase transition for different training strategies. The horizontal line indicates the statistical uncertainty of the SSE reference
    energy, and the vertical dashed line indicates the critical
    detuning. PQMC simulations are performed with $N_w=6000$
walkers.}
\end{figure}

Figure~\ref{fig:additional_energy} compares the ground-state energy
obtained with these approaches across the range of detunings considered
in the main text. Applying PQMC to the variationally optimized RBM
substantially improves upon VMC alone, demonstrating the benefit of the
projective step even when starting from an optimized variational state.
The accuracy of PQMC+VMC, however, deteriorates close to the transition and within the checkerboard phase. This behavior is likely due to an increasing effect of the finite walker population that impedes the suppression of the bias in VMC simulations. PQMC+Data shows a similar dependence, although
experimental pretraining provides a particularly accurate initialization
deep in the checkerboard phase. In contrast, sPQMC+Data maintains a
more uniform accuracy across the different regimes because the initial
data-trained guiding wavefunction is further refined using the projected walker
distribution.

\begin{figure}
    \centering
    \includegraphics[width=\columnwidth]{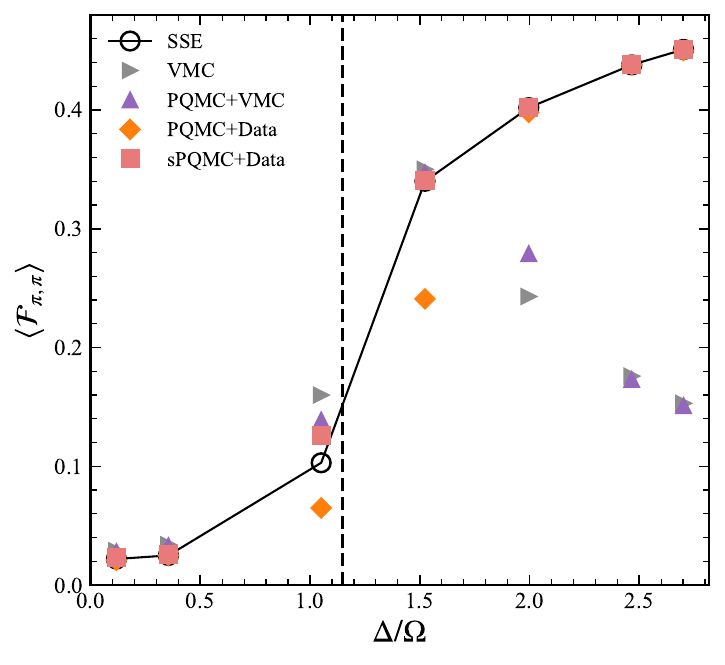}
    \caption{\label{fig:additional_magnetization}
    \textbf{Checkerboard order parameter for different
    guiding-wave-function optimization strategies.}
    Checkerboard order parameter as a function of the detuning
    $\Delta/\Omega$. The vertical dashed line indicates the critical
    detuning. }
\end{figure}

The same behavior is reflected for the checkerboard order parameter,
shown in Fig.~\ref{fig:additional_magnetization}. While all methods
provide similar results away from the ordered regime, differences
become more pronounced around and beyond the transition. In particular,
the deterioration of the fixed VMC-optimized RBM is inherited by
PQMC+VMC at large detuning, whereas the experimentally initialized
methods more closely reproduce the SSE results. The agreement of
sPQMC+Data across the full range further illustrates the advantage of
combining a physically informed initialization with the subsequent
self-learning refinement of the guiding wavefunctions.

In principle, a variationally optimized guiding wavefunction could
also be used to initialize the self-learning procedure (sPQMC+VMC).
This would, however, require a separate VMC ground-state optimization
before the projective simulation. Experimental pretraining is
qualitatively different in this respect: it amounts to
likelihood-based learning from an already available set of measured
configurations and does not require variational energy minimization. Once this comparatively inexpensive initialization is performed, sPQMC uses the projection itself both to
estimate ground-state properties and to generate the configurations
used to improve its guiding wavefunction. 

Experimental information can additionally be used to initialize the
PQMC walker population. For the RBM guiding wavefunction, we have found, though it was not shown in the manuscript, that initializing
the walkers from experimental configurations accelerates convergence
relative to a random initial walker distribution. By contrast, no
appreciable improvement was observed for its RNN counterpart. This is likely because the RNNs
provide a substantially more accurate representation of the target
distribution. This indicates that data-informed walker initialization
is likely most beneficial when the initial guiding wavefunction is less
accurate.

\subsection{Off-diagonal observables}
\label{app:offdiag}
We further assess the ability of data-enhanced self-learning PQMC to
estimate off-diagonal observables by considering the expectation value
of the Pauli-$X$ operator~\cite{moss24},
\begin{equation}
\label{eq:expval_x}
    \expval{\hat{\sigma}^x}
    =
    \frac{1}{N_w}
    \sum_{\boldsymbol{x}}
    \frac{1}{N}
    \sum_i
    \frac{\psi_G(\boldsymbol{x}'_i)}
         {\psi_G(\boldsymbol{x})},
\end{equation}
where $\boldsymbol{x}'_i$ denotes the configuration obtained from
$\boldsymbol{x}$ by flipping spin $i$. Unlike diagonal observables,
this estimator explicitly depends on ratios of guiding wavefunction
amplitudes between different configurations and is therefore
particularly sensitive to the quality of the learned guiding wavefunction~\cite{perez2026unbiaseddiffusionmontecarlo}. Note that, $N_w$ in Eq. \ref{eq:expval_x} represents the number of walkers accumulated during the forward walking procedure, thus, $\expval{\hat{\sigma}^x}$ renders a pure estimator.

\begin{figure}
  \centering
  \includegraphics[width=\columnwidth]{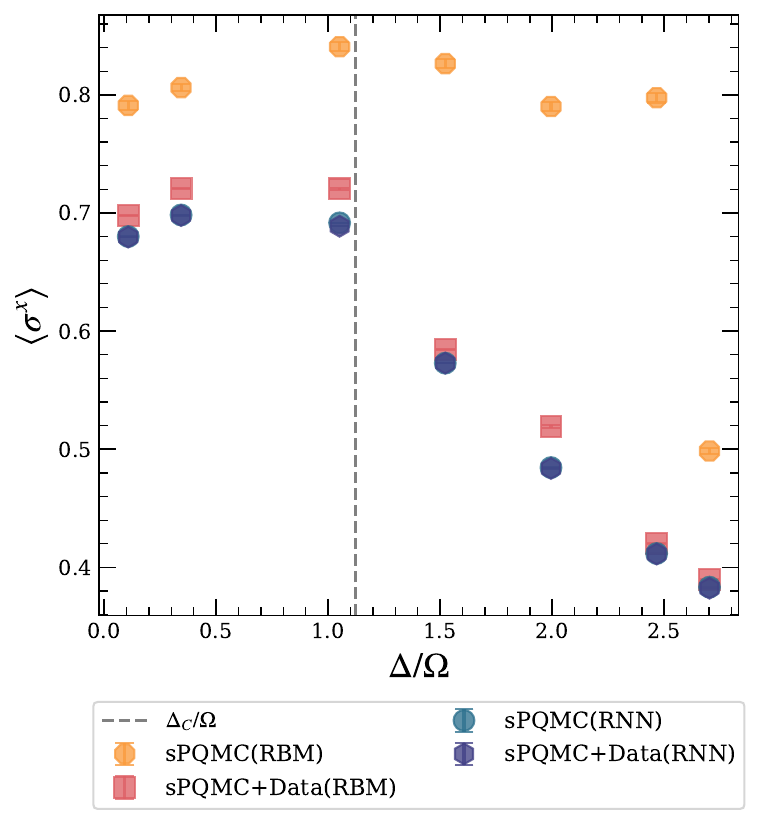}
  \caption{\label{fig:x_phase_diagram} \textbf{Off-diagonal observable across the phase transition.}
Expectation value of the Pauli-$X$ operator (see (Eq. \ref{eq:expval_x}))
as a function of the detuning $\Delta/\Omega$, obtained from sPQMC
using the forward-walking technique. Data are collected using the same
simulation protocol as in Figs.~\ref{fig:relative_error_phase_diagram} and
\ref{fig:phase_diagram}.}
\end{figure}

Figure~\ref{fig:x_phase_diagram} shows
$\langle\hat{\sigma}^x\rangle$ as a function of the detuning
$\Delta/\Omega$. The suppression of $\langle\hat{\sigma}^x\rangle$
upon entering the checkerboard regime is captured by the
sPQMC calculations with RNNs, as well as by the data-enhanced
RBM-guided calculation. The results are consistent with the VMC
calculations of Ref.~\cite{moss24}. These results provide an additional
test of the learned guiding wavefunctions beyond the diagonal
observables considered in the main text. In particular, the RNN guiding wavefunction
accurately captures the amplitude ratios entering the off-diagonal
estimator, while experimental pretraining substantially improves the
less expressive RBM guiding wavefunction.

\subsection{Convergence of walker population}
\label{app:walker_convergence}

\begin{figure}
    \centering
    \includegraphics[width=\columnwidth]{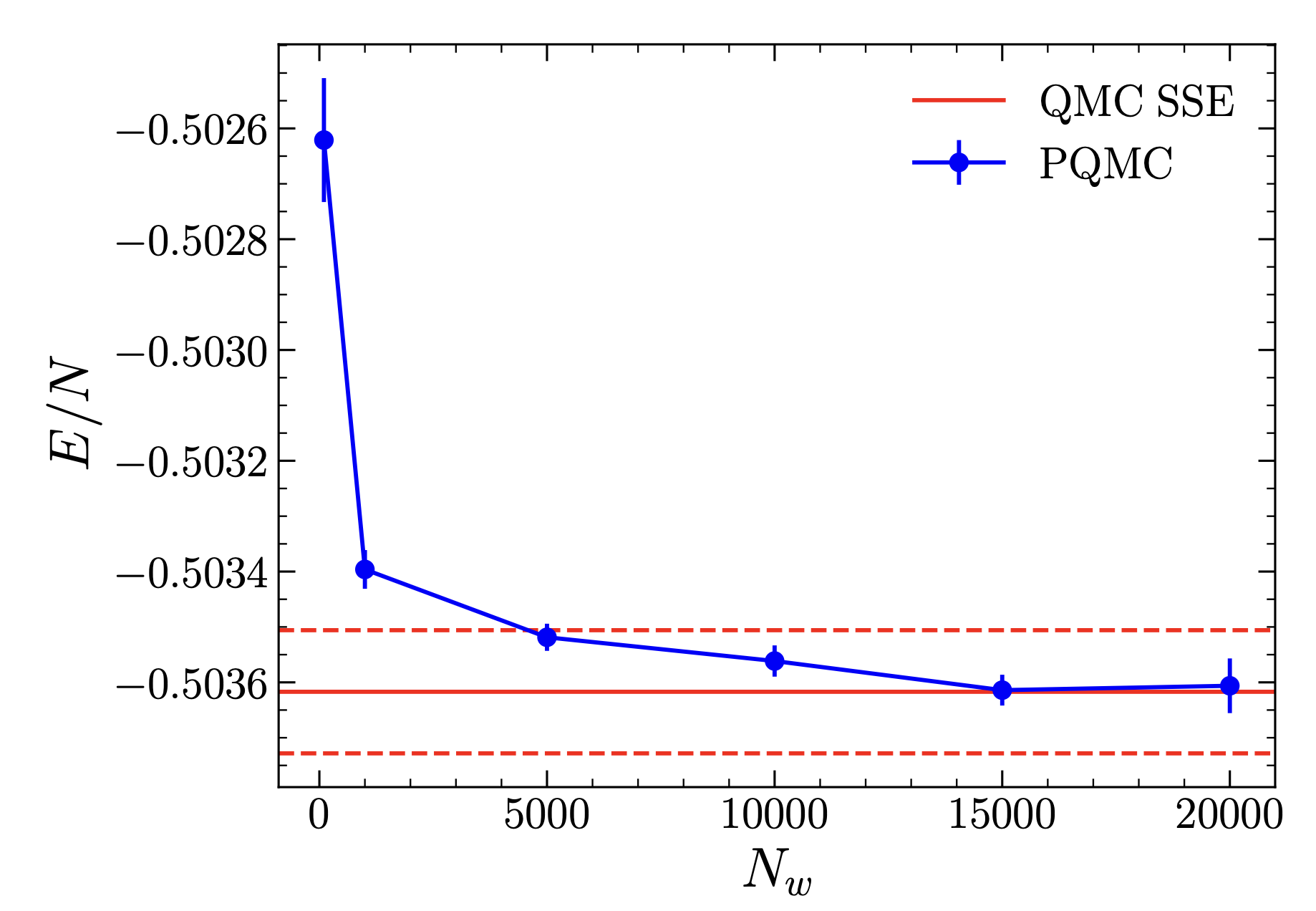}
    \caption{\label{fig:walker_convergence}
    \textbf{Convergence of PQMC with walker population.}
    Ground-state energy per site $E/N$ obtained with PQMC using a fixed
    RBM guiding wavefunction as a function of the walker population
    $N_w$. The guiding wavefunction is obtained from a separate
    self-learning optimization. The solid red line indicates the
    SSE reference energy, while the
    dashed red lines indicate its statistical uncertainty. Simulations are performed at $\Delta/\Omega=1.051$.}
\end{figure}

We examine the dependence of the PQMC estimates on the walker
population $N_w$. Figure~\ref{fig:walker_convergence} shows the
ground-state energy per site at $\Delta/\Omega = 1.051$ obtained with PQMC using a fixed RBM
guiding wavefunction previously optimized in a separate self-learning
simulation. As $N_w$ is increased, the estimated energy systematically
approaches the SSE reference value. This behavior confirms that the
residual deviations at finite $N_w$ originate from finite-population
effects and can be systematically reduced by increasing the walker
population.

This result also clarifies the role of self-learning in the calculations
presented in the main text. A sufficiently accurate fixed guiding wavefunction can in
principle yield converged PQMC results when combined with a large enough
walker population. Self-learning instead improves the guiding wavefunction during the
projection, reducing the sensitivity of the calculation to the initial
wavefunction and allowing accurate estimates to be obtained with
smaller walker populations. Experimental pretraining further improves
this starting point, particularly for less expressive guiding wavefunctions such as the RBM.

\subsection{Binder cumulants}
\label{app:ffs_binder}
To further assess the reliability of sPQMC with RNN guiding wavefunctions, we compute the Binder cumulant of the checkerboard order parameters used in the production of Fig.~\ref{fig:fss} and its finite-size scaling. The Binder cumulant of the checkerboard order parameter is defined as:
\begin{equation}
    \label{eq:binder}
    U_4=
    \frac{3}{2}\left[
    1-\frac{\langle\mathcal{F}_{\pi,\pi}^{4}\rangle}
    {3\langle\mathcal{F}_{\pi,\pi}^{2}\rangle^{2}}
    \right].
\end{equation}
Near the critical point, the quantity obeys the finite-size scaling ansatz
$
    U_4=f\left[
    \frac{\Delta-\Delta_c}{\Omega}L^{1/\nu}
    \right],
$
where $\Delta_c$ is the critical detuning, $\nu$ is the correlation-length exponent, and $f$ is a scaling function. At criticality, $U_4$ is approximately independent of the system size, up to finite-size corrections.

\begin{figure}
    \centering
    \includegraphics[width=\columnwidth]{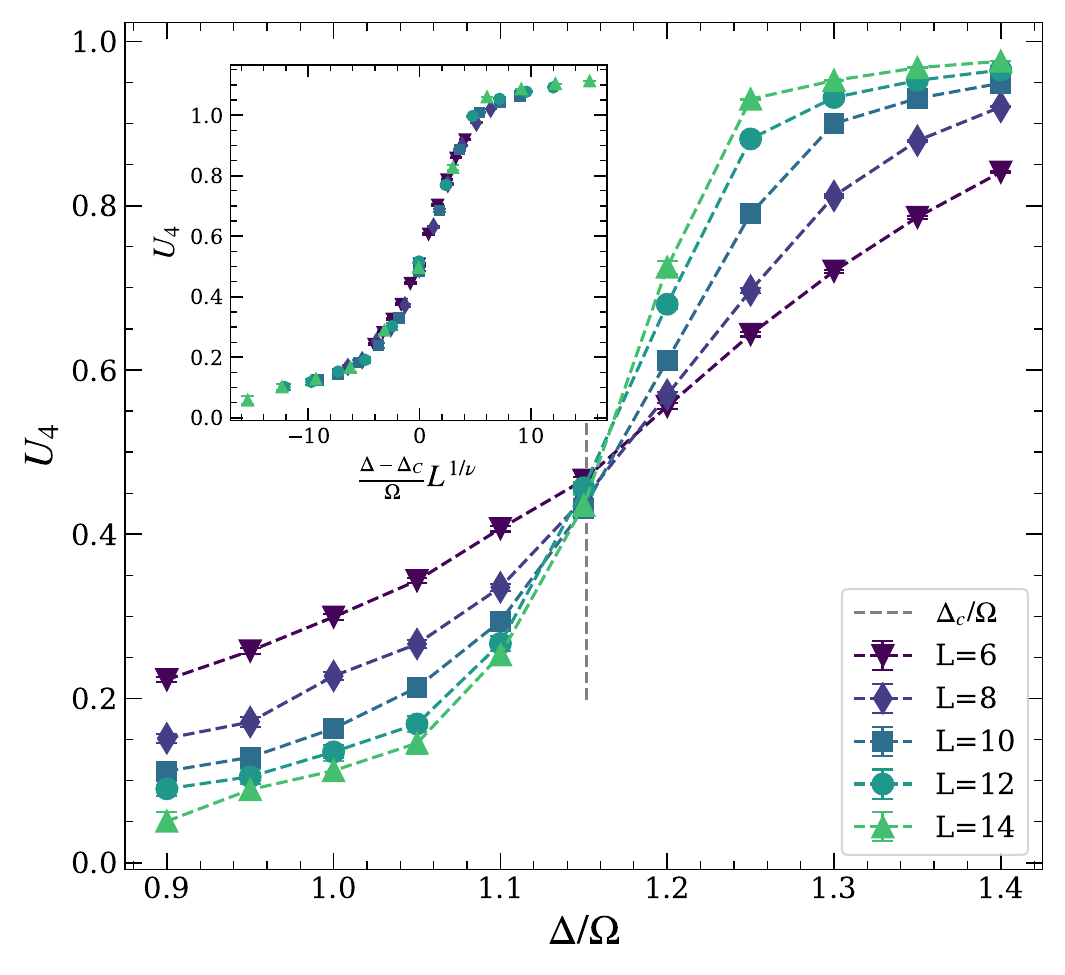}
    \caption{
    \textbf{Binder cumulant across the disordered-to-checkerboard phase transition.}
    Binder cumulant of the checkerboard order parameter as a function of the detuning $\Delta/\Omega$ for different $L\times L$ lattices at $R_b/a=1.2$, obtained from sPQMC simulations with RNN guiding wavefunctions without experimental-data pretraining. The inset shows the corresponding data collapse of the Binder cumulant. 
    }
    \label{fig:binder_cumulant}
\end{figure}

As shown in Fig.~\ref{fig:binder_cumulant}, the Binder cumulant curves for different system sizes cross near a common critical point, as expected. A finite-size scaling collapse yields $\Delta_c/\Omega\approx1.151(1)$ and $\nu\approx0.641(5)$, consistent with the results reported in the main text. These results support the reliability of sPQMC with RNN guiding wavefunctions in locating and characterizing the phase transition in the Rydberg array.

\section{Experimental parameters}
\label{app:experimental}

Following the organization used in Ref.~\cite{moss24}, Table~\ref{tab:exp} collects the experimental and data-processing quantities used in this work. The dataset consists of projective fluorescence measurements of the final spin configuration after the detuning is swept from a large negative value toward the checkerboard regime. In order to deal with the limited availability of projective measurements, we combine data from the seven sweep rates available. It yields 4000 configurations per detuning, which we further augmented leveraging the symmetries of the Rydberg Hamiltonian on the square lattice by $\pi/2$, $\pi$, and $3\pi/2$ rotations. In total, we obtain 16000 configurations per detuning for NQS pretraining.

\begin{table}[H]
\caption{Experimental Rydberg-array and dataset parameters.}
\scriptsize
\label{tab:exp}
\begin{ruledtabular}
\begin{tabular}{ll}
Quantity & Value \\
\hline
Atomic species & $^{87}$Rb \\
Array geometry & square, OBC \\
Main system size & $16\times16$ ($N=256$) \\
Lattice spacing & $a\simeq6.7~\mu$m \\
Blockade ratio & $R_b/a\simeq1.15$ \\
Interactions cutoff & $R_0=4a$ \\
Rabi frequency & $\Omega=2\pi\times4.3$~MHz \\
Sweep rates & 15, 21, 30, 42, 60, 85, \\
& and 120~MHz/$\mu$s \\
Snapshots / sweep / detuning & 500 (1000 at 15~MHz/$\mu$s) \\
Combined snapshots per detuning & 4000 \\
Data augmentation & rotations by $\pi/2$, $\pi$, $3\pi/2$ \\
Training set / detuning & 16000 \\
Reported readout fidelity & $\sim99\%$ \\
\end{tabular}
\end{ruledtabular}
\end{table}


\section{Guiding wavefunction parameters}
\label{app:nqsparams}

Table~\ref{tab:nqs} lists the neural guiding wavefunction settings. The RNN is a patched architecture following Ref.~\cite{sprague24}.

\begin{table}[!htbp]
\caption{Neural guiding wavefunction parameters used in the self-learning and data-enhanced self-learning PQMC calculations.}
\scriptsize
\label{tab:nqs}
\begin{ruledtabular}
\begin{tabular}{lll}
Parameter & RBM & Patched RNN \\
\hline
Visible sites & 256 & 256 \\
Hidden size & $N_h=512$ & 32 \\
Hidden density & $N_h/N=2$ & --- \\
Recurrent layers & --- & 1 \\
Trainable parameters & 131,840 & 4,176 \\
Patch size & --- & $2 \times 2$ \\
Data-pretraining loss & KL  & KL  \\
Data optimizer & CD-$k$ \& SGD with momentum & Adam \\
\end{tabular}
\end{ruledtabular}
\end{table}


\section{sPQMC simulation parameters}
\label{app:pqmcparams}

Table~\ref{tab:pqmc} lists the settings of the projective step of the sPQMC method used in the calculations reported in the main text.

\begin{table}[H]
\caption{sPQMC simulation parameters.}
\scriptsize
\label{tab:pqmc}
\begin{ruledtabular}
\begin{tabular}{ll}
Quantity & Value \\
\hline
Projector & continuous-time PQMC \\
Population control & dynamic $E_{\rm T}$ with hard $N_w$ boundary \\
Target random walker population & $6000$ (Figs.~\ref{fig:relative_error_phase_diagram}, \ref{fig:phase_diagram}, \ref{fig:x_phase_diagram})\\
& $4096$ (Fig.~\ref{fig:relative_error_convergence}, ~\ref{fig:fss}) \\
Self-learning training samples & $\sim 10^5$ (Figs.~\ref{fig:energy_vs_walker}, \ref{fig:relative_error_phase_diagram}, \ref{fig:phase_diagram}, \ref{fig:fss}) \\
& $\sim2\times10^3$ (Fig. \ref{fig:relative_error_convergence}) \\
& $\sim1.6 \times 10^{4}$ (Fig.~\ref{fig:system_size_scalings})\\
Imaginary timestep & $2 \times 10^{-2}$ \\
Projection iterations & $4\times 10^4$ (Figs.~\ref{fig:energy_vs_walker}, \ref{fig:relative_error_phase_diagram}, \ref{fig:phase_diagram}) \\
& $1.2\times 10^4$ (Fig. \ref{fig:relative_error_convergence})\\
&$3\times 10^{4}$ Fig.~(\ref{fig:system_size_scalings}, \ref{fig:fss}, \ref{fig:binder_cumulant}) \\
Self-learning training iterations & $\sim10^4$ (Figs.~\ref{fig:energy_vs_walker}, \ref{fig:relative_error_phase_diagram}, \ref{fig:phase_diagram}, \ref{fig:fss})\\
& $\sim10^3$ (Fig. \ref{fig:relative_error_convergence}, \ref{fig:system_size_scalings})\\
Forward walking & 1000 iterations \\
\end{tabular}
\end{ruledtabular}
\end{table}


\section{Execution time}
\label{app:exec_time}

Table~\ref{tab:exe_time} lists the execution time per step (in seconds) used to produce Fig.~\ref{fig:relative_error_phase_diagram}, Fig.~\ref{fig:phase_diagram}, and Fig.~\ref{fig:x_phase_diagram}. Note that the reported times are only in the specific setting of our code, and they could be improved upon with further optimization.

\begin{table}[h]
\caption{Execution time per step of procedures used in the data-enhanced sPQMC simulations}
\scriptsize
\label{tab:exe_time}
\begin{ruledtabular}
\begin{tabular}{lll}
Procedure & sPQMC (RBM) &sPQMC(RNN) \\
\hline
Device & AMD EPYC 7543 & NVIDIA H200\\
Projection & $\sim 3.44\times10^{-1}$ s & $\sim 7.00 \times 10^{-1}$ s\\
Training (Experimental data) &  $\sim  8.59 \times 10^{-2}$ s & $\sim2.40\times 10^{-3}$ s\\
Training (Self-learning) & $\sim  9.94 \times 10^{-2}$ s & $\sim 3.42 \times 10^{-3}$ s \\
\end{tabular}
\end{ruledtabular}
\end{table}

\newpage


\section{Curves collapse parameters}
\label{app:scaling_collapse}

Table~\ref{tab:fss} lists the simulation details in producing curves in Fig.~\ref{fig:fss} and Fig.~\ref{fig:binder_cumulant} along with the fitting parameters used in the finite size scaling analysis.

\begin{table}[h]
\caption{Simulation details and curves collapsing parameters used to determine disordered-to-checkerboard phase transition critical exponents}
\scriptsize
\label{tab:fss}
\begin{ruledtabular}
\begin{tabular}{ll}
Quantity & value \\
\hline
Finite-size scaling sizes & $L= 6,8,10,12,14$ ($R_b/a\simeq1.2$)\\
& $L= 8, 10,12,14,16$ ($R_b/a\simeq1.15$)\\
Finite-size scaling detuning & $\Delta/\Omega\simeq0.9$--$1.4$ \\
Optimizer & Nelder-Mead \\
Polynomial order & 6 ($R_b/a\simeq1.2$)\\
& 9 ($R_b/a\simeq1.15$)\\
Search boundary for $\Delta_c/\Omega$ & $\left[0.9, 1.4\right]$ \\
Search boundary for $\beta/\nu$ & $\left[0.0, 1.0\right]$ \\
Search boundary for $1/\nu$ & $\left[1.0, 2.0\right]$
\end{tabular}
\end{ruledtabular}
\end{table}
\end{document}